\documentclass[12pt,fleqn]{article}
\usepackage{t1enc}
\usepackage[latin1]{inputenc}
\usepackage[dvips]{graphicx}
\usepackage{flafter}
\usepackage[dvips]{rotating}
\usepackage{latexsym}
\newcounter{saveeqn}

\begin{document}
\begin{center}
\Large {\bf Frequency dependent effective conductivity of
two-dimensional metal-dielectric composites }
\end{center}
\vspace{0.3cm}

\begin{center}
LOTFI ZEKRI\footnote
{Permanent address:\em U.S.T.O., D\'{e}partement de Physique, L.E.P.M., B.P.1505 El M'Naouar, Oran, Algeria. Email: lzekri@yahoo.com}

{\em Institute for Theoretical Physics, Cologne University, D-50923 K\"oln, Germany
E-mail: zekri@thp.uni-koeln.de}
\end{center}

\vspace{0.5cm} \centerline{\bf Abstract}

We analyze a random resistor-inductor-capacitor $(RLC)$ lattice model of 2 dimensional metal-insulator composites. The results are compared with Bruggeman's and Landauer's Effective Medium Approximations predictions where a discrepancy was observed for some frequency zones. Such a discrepancy is mainly caused by the strong conductivity fluctuations. Indeed, a two-branches distribution is observed for low frequencies.
We show also that at $p_c$ the so-called Drude peak vanish by increasing the system size and increase for vanishing losses.

\vspace{0.5cm}

\section{Introduction
\label{sec_01}}

The optical properties of composite materials are very different from those of their constituents particularly in metal insulator composites~\cite{01,02}. Theses differences are mainly observed at percolation threshold $p_c$ at which one of the two components of the composite first forms a connected path extending through the system~\cite{03}. Bruggeman's and Landauer's Effective Medium Aproximation (BEMA~\cite{04}, LEMA~\cite{05}) provides the most useful way to describe these properties.

Zeng et al.~\cite{06}, following the work of Koss and Stroud~\cite{07}, analyzed the optical properties of a two-dimensional normal-metal-insulator composite.

They found a surface plasmon resonance peak of the real part of the conductivity for various concentrations above and below $p_c$ in agreement with LEMA. For $dc$ case, Watson and Leath~\cite{08} gave equation to correct the slope of the curve at the critical concentration. Koss and Stroud generalized this equation to their $ac$ problem.
 The so-called surface plasmon modes are known to be an important absorption mechanism in random metal-insulator composites~\cite{09}. Drude peak, corresponding to nonzero dc conductivity at and above $p_c$ was also found. This peak is due at $p_c$ to the finite size effects and is expected by Koss and Stroud to vanish in the limit of infinite systems.
 For system size $100\times 100$ Zeng chose a characteristic relaxation time value of 10 in $RLC$ systems corresponding to a large loss in the metallic component $R={10}^{-1}$ ($\tau=L/R$).
 For vanishing losses the local field and the $ac$ conductivity fluctuations are strong mainly at percolation threshold $p_c$~\cite{10,11}. Thus calculations for only 5 samples, as done by Zeng, for vanishing losses, will  not be statistically sufficient.

Zeng et al. used Frank and Lobb algorithm~\cite{12}, for frequencies ranged from ${10}^{-2}$ to 1 (expressed in units of the plasmon frequency $\omega_p$). This algorithm is very fast but based on many division operations, which causes a numerical instabilities both for smaller frequencies and losses. In order to compute the effective conductivity for the corresponding $2d$ resistor network for smaller frequencies and losses, we use a method of exactly solving Kirchoff's equations, extensively described in Refs~\cite{13}. By using this method the current is preserved in all nodes of the square lattice and the total entering and outgoing currents of the sample are accurately identical.

In this paper we present a numerical study on $2d$ resistor network model for a large frequency range including the plasmon frequency $\omega_p$ around the bond percolation threshold $p_c$ for smaller loss and larger size.

 Confirming Koss and Stroud prediction we show that the Drude peak decreases by increasing the system size. We show also that for vanishing losses Drude peak increase and, at percolation threshold $p_c$, for low frequencies the distribution of the real part of the effective conductivity is splitted in two branches, the first one is due to the non percolate samples and the second branch to the percolate ones. The results allow us to test the validity of the above cited BEMA and LEMA.
\section{Model description
\label{sec_02}}
It is well known that in the optical frequency region a semi-continuous metallic film can be modelled as a $2d$ $RL-C$ lattice~\cite{02}.
Capacitors $C$ stand for the dielectric grains that have dielectric constant $\epsilon_d$, while the inductance $L$ represents the metallic grains characterized for Drude metal by the following dielectric function~\cite{14},
\begin{eqnarray}
\epsilon_m(\omega)=\epsilon_b-(\omega_p/\omega)^{2}/[1+i\omega_{\tau}/\omega]
\end{eqnarray}
where $\epsilon_b$ is the contribution to $\epsilon_m$ due to the inter-band transitions, $\omega_\tau=1/\tau$ is the relaxation rate and $\omega_p$ is the plasmon frequency.
While for dielectric grains, the dielectric function is,
\begin{eqnarray}
\epsilon_d=cte
\end{eqnarray}

The metal conductivity $\sigma_m=-i\epsilon_m\omega/4\pi$ is characterized by positive imaginary part in the optical frequency region (inductive behavior)~\cite{02}. We can then model the metallic grains as inductances $L$ with loss $R$ ($\sigma_m=(R+iL\omega)^{-1}$) while the dielectric grains can be represented by capacitances $C$ ($\sigma_d=iC\omega$).
We can take without loss of generality, $L=C=\omega_p=1$, ($\omega_p$ corresponds to the case where
$\mid\sigma_m\mid\simeq\mid\sigma_d\mid$ for vanishing losses, i.e., $C\omega_p=1/L\omega_p$).
The inductance $L$ and the capacitance $C$ can be assumed to be constant in the range between $1/\tau$ and $\omega_p$, we express in this paper the frequency  in units of the plasmon frequency ($\tilde{\omega}=\omega/\omega_p<1$).

We model the metal dielectric composite films as a square lattice composed of metallic bonds with conductivity :
\begin{eqnarray}
\sigma_m=(R+i\tilde{\omega})^{-1}
\end{eqnarray}
and concentration $p$, and dielectric bonds with conductivity:
\begin{eqnarray}
\sigma_d=i\tilde{\omega}
\end{eqnarray}
and concentration $1-p$.

In order to compare with Zeng's results, we model the metal by a capacitance parallel to $R L$ branch, this is called $R L\|C$ model to be distinguished from the $R L$ one.

\section{Results and discussion
\label{sec_03}}
We consider a lattice of size $n\times n$ ($n=100$, $200$ and $600$). If we choose $R={10}^{-3}$, the smaller system size is larger than the correlation length (following Zekri et al.~\cite{11}, the correlation length behaves as $R^{-1/2}$). We choose then losses $R$ not smaller than the above value so that we ensure a correlation length always smaller than the system size. This avoids the size dependent effects of the conductivity.
The effective conductivity here is averaged from 550 samples and compared to Zeng's simulations for only 5 samples.

Let us first reproduce Zeng's results on the effective conductivity by exactly solving Kirchoff laws  ~\cite{13} for an $RL\|C$ model. We compare them to $RL$ model at $p_c$ with $R={10}^{-1}$. In $Figure 1a$, the two models seem to produce identical results except near $\tilde{\omega}=1$, where the real part of the conductivity seems to vanish for $RL\|C$ model while it saturates for $RL$ model,
the divergence between the two models start from the frequency $\tilde{\omega}=0.35$. Large fluctuations of this conductivity are observed below $\tilde{\omega}={10}^{-2}$. We show also that the two models are in a good agreement with BEMA for all the frequency range investigated and are different from LEMA.
Comparison between the two models at $p_c$ with $R={10}^{-3}$ is done for sizes 100 and 200. They produce also identical results but the fluctuations extends to the whole frequency range investigated. Then we have done calculations for large samples using $RL$ model. In $Figure 1b$ comparison between $RL$ model for system sizes 100, 200 and 600 and both BEMA and LEMA is done at $p_c$ with $R={10}^{-3}$. The size effect on Drude peak is also investigated. The results obtained and BEMA are alike only for frequencies ranged from ${10}^{-2}$ to 1, and they follow qualitatively LEMA for low frequencies, where Drude peak increase by decreasing the loss. We see also in this figure that Drude peak decreases by increasing the system size and should vanish in the limit of infinite systems as predicted by Koss and Stroud~\cite{07}. The data in the described curves are directly averaged (i.~e.~Arithmetic Average (AA) of the 550 samples for sizes 100 and 200 and 30 samples for the size 600). In order to check the validity of this average, the same results for the size 200 are plotted using Logarithmic Average(LA). The LA curve and the AA one are alike for frequencies ranged from ${10}^{-2}$ to 1, and they diverge for low frequencies.

It is then important to investigate the conductivity distributions for frequencies in the range examined here.
For $\tilde{\omega}=1$ the distribution of the real part is Gaussian centered at 1. For $\tilde{\omega}={10}^{-1}$ the distribution becomes broader and at $\tilde{\omega}={10}^{-2}$ it becomes poissonian (the distributions for $\tilde{\omega}=1$, ${10}^{-1}$ and ${10}^{-2}$ are not plotted here). In $Figures 2a$ we show for $\tilde{\omega}={10}^{-3}$ that the distribution is splitted in two branches, the first one has the form of a peak centered at the vicinity of zero and the second one has a low height and is broad, its width tends towards large positive values. The two branches distribution observed for frequencies less than ${10}^{-2}$ are the reason of the large divergence between AA and LA. In $Figures 2b$ and $2c$ we compare for the frequency $\tilde{\omega}={10}^{-3}$ the distributions of the real part for concentrations $p$ of the metallic grains ($RL$ bound) below and above $p_c$. Above $p_c$ means $p-p_c =1/N^{1 / \nu}$ and below $p_c$, $p-p_c =-1/N^{1 / \nu}$ where the connectivity length exponent $\nu$~\cite{03} is $4/3$ for $d=2$ and $N$ is the system size. The peak centered in the vicinity of zero (first branch) increases below $p_c$ and tends to disappear above $p_c$, whereas the width of the second branch tends to disappear below $p_c$, and increase above $p_c$ and becomes dominant. Therefore the two branches appear only at $p_c$.
The first branch is related to the non percolate samples and the second branch is related to the percolate ones. Similar behavior of this conductivity distribution has been observed recently~\cite{15} for the distribution of the critical links (singly connected links~\cite{03,16}). The disappearance of one branch corresponds to that observed for critical links below $p_c$ (also observed when the system size increase~\cite{15}). Such branch corresponds to configurations for which the infinite cluster has a very small number of critical links.

\section{Conclusion
\label{sec_04}}
In this paper we have compared ($RLC$) lattice model with BEMA and LEMA at and near the percolation threshold $p_c$ for large and small loss in the metallic component.
For large loss the results and BEMA are alike for all the frequency range investigated, but for small loss they are alike only for frequencies ranged from ${10}^{-2}$ to 1, and they follow qualitatively Landauer's approximation for lower frequencies.
We have also shown that for small loss the arithmetic and the logarithmic average of the real part of the effective conductivity diverge for frequencies less than ${10}^{-2}$ due to two branches distribution of the real part of the effective conductivity. The first one is due to non percolate samples and the second branch is due to percolate ones.
We showed also in this paper that Drude peak increases by decreasing the metallic loss and vanish with the size as predicted by Stroud~\cite{07} at $p_c$.


\vspace{0.5cm}
{\bfseries Acknowledgments}

\vspace{0.2cm}
I wish to thank Professors D.Stauffer and N.Zekri for helpful discussions and DAAD for financial support during the progress of this work.

\clearpage

\begin{figure}[h]
\includegraphics[width=7.0cm,height=15.0cm, angle=-90]{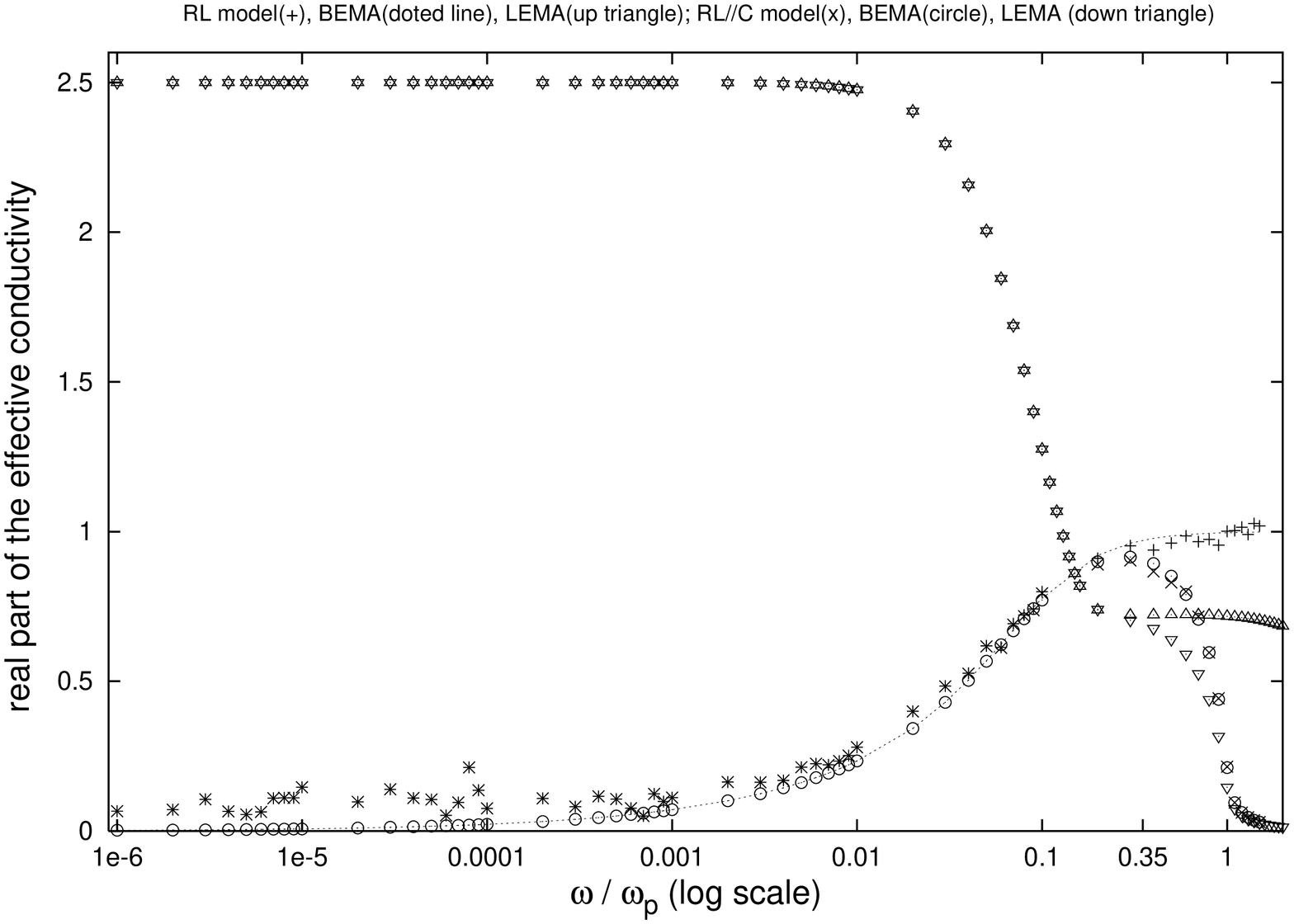}
\includegraphics[width=7.0cm,height=15.0cm, angle=-90]{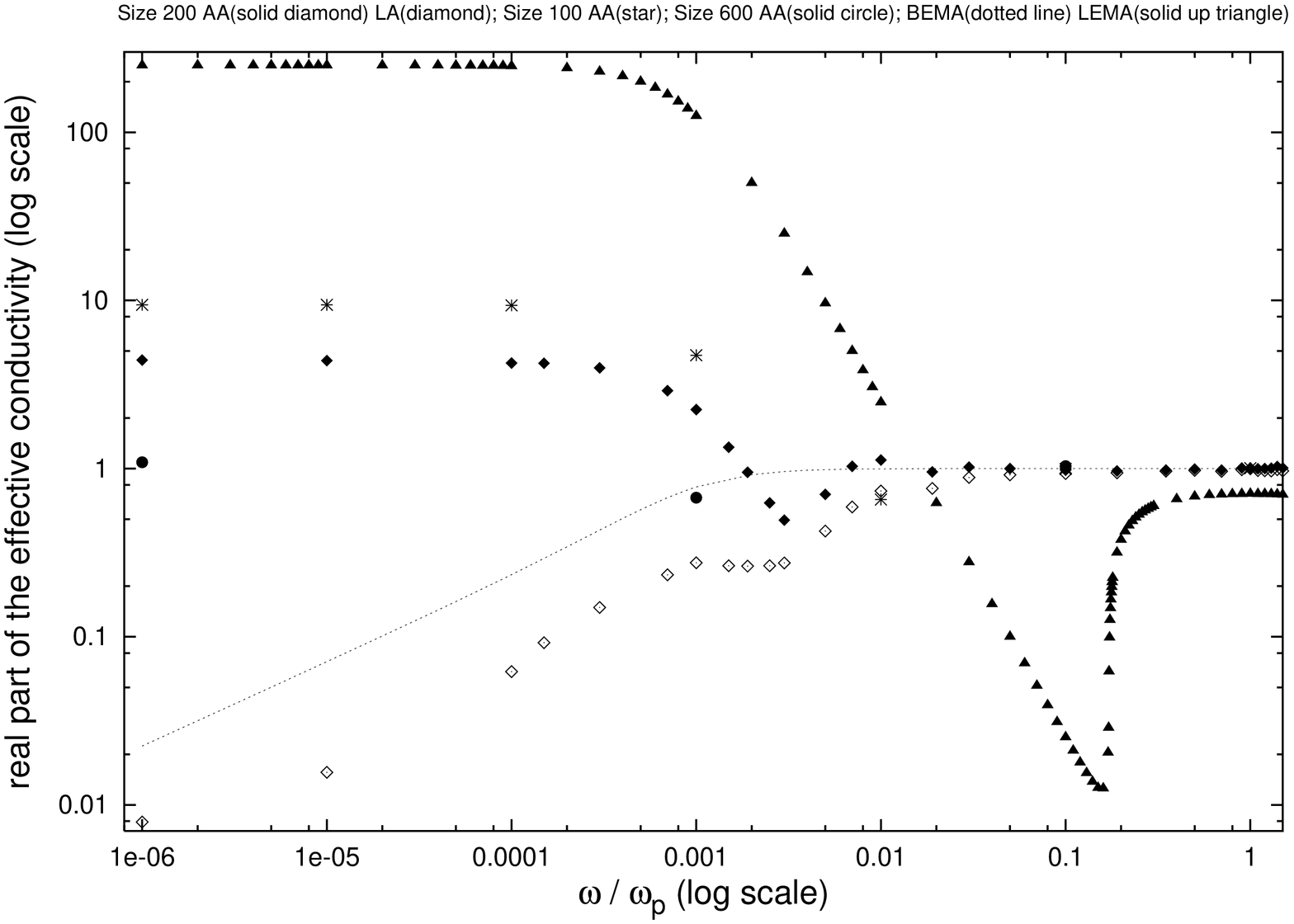}
\caption { $a)$ $R={10}^{-1}$, Arithmetic Average (AA) over 5 samples of size 100. $b)$ $R={10}^{-3}$, $RL$ model, AA over 550 samples for the sizes 100 and 200 and 30 samples for the size 600. Logarithmic Average(LA) for the size 200. The real part of the effective conductivity is expressed in arbitrary units and the frequency is expressed in units of the plasmon frequency.}
\end{figure}

\clearpage

\begin{figure}[h]
\includegraphics[width=7.0cm,height=15.0cm, angle=-90]{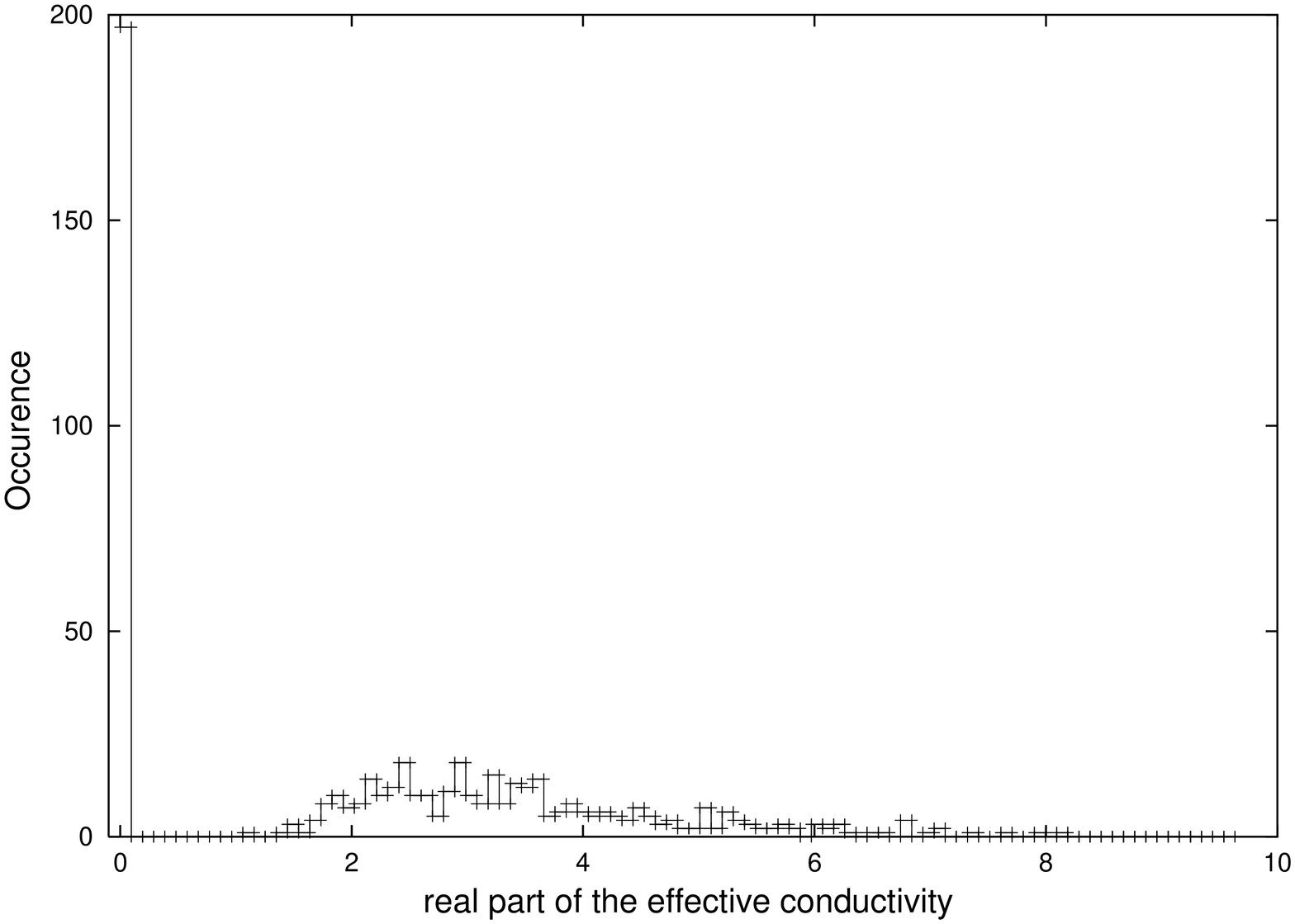}
\includegraphics[width=7.0cm,height=15.0cm, angle=-90]{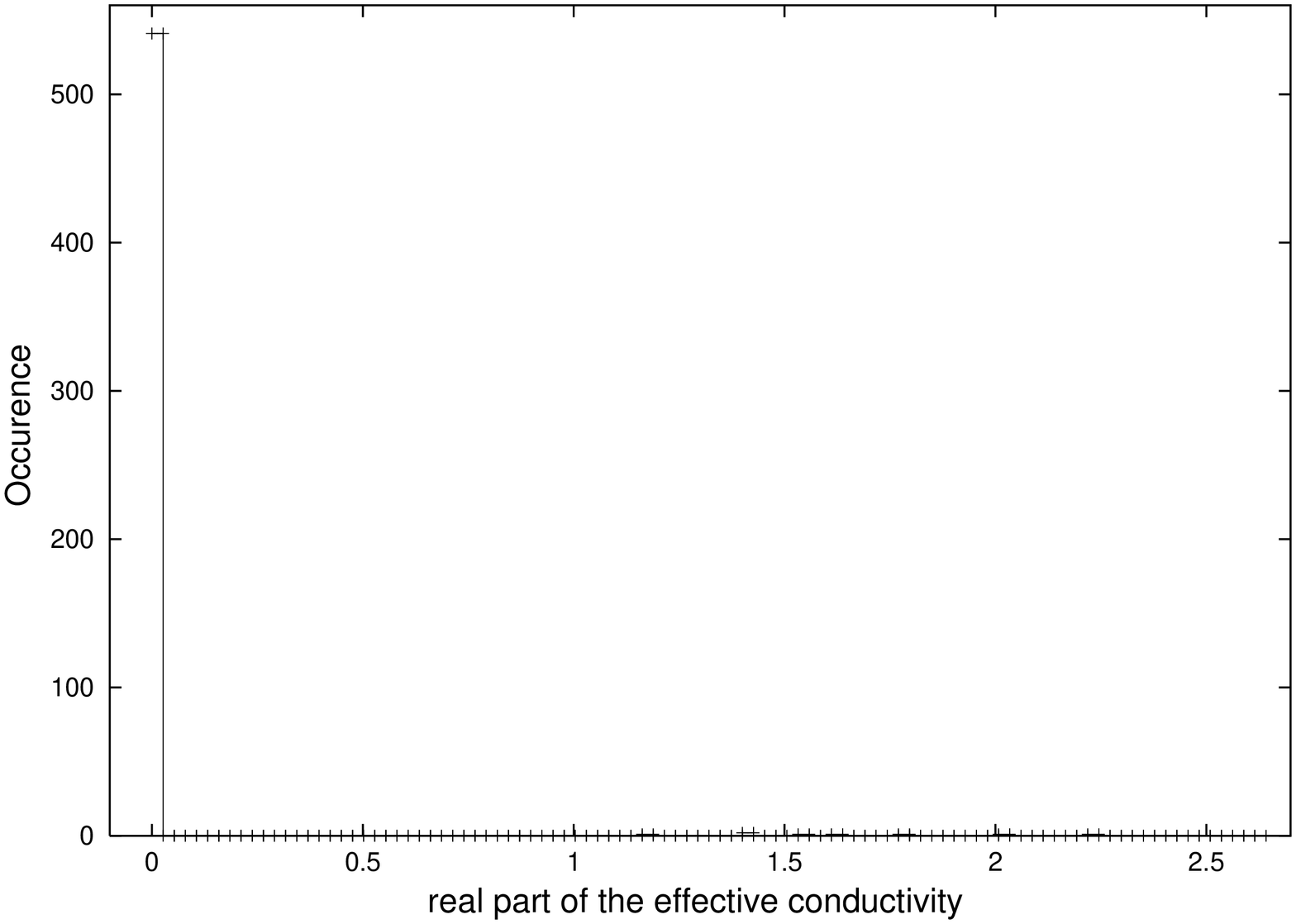}
\includegraphics[width=7.0 cm,height=15.0cm, angle=-90]{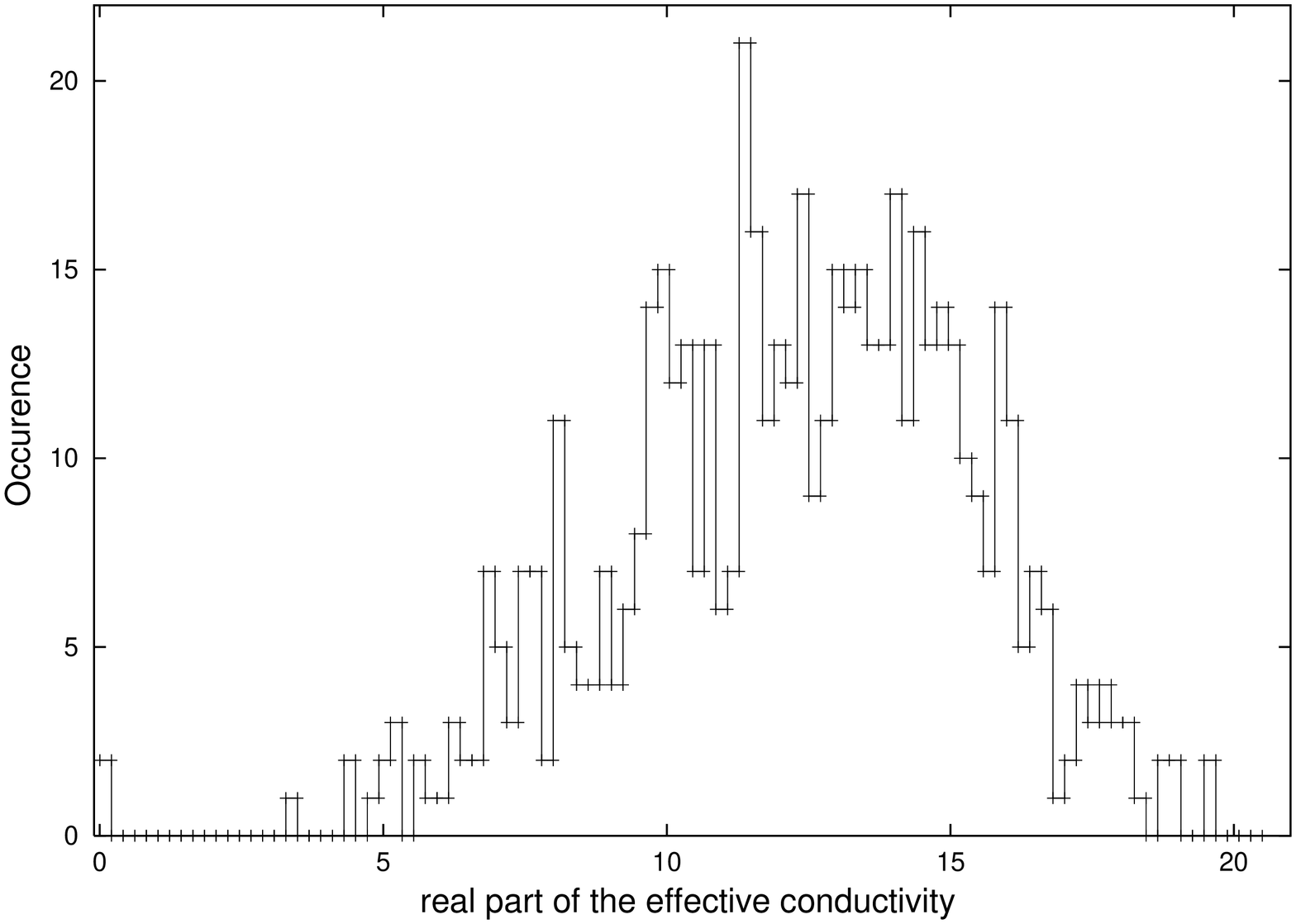}
\caption {Distribution of the real part of the effective conductivity expressed in arbitrary units, using $R L$ model and AA over 550 samples of size 200 at frequency $\tilde{\omega}={10}^{-3}$, at $p_c$ $(a)$, below $p_c$ $(b)$ and above $p_c$ $(c)$}
\end{figure}

\clearpage
\noindent
\vspace{1cm}


\begin{thebibliography}{99}
\markboth{Bibliography}{}
\bibitem{01} Landauer R., in {\it Electrical Transport and Optical Properties of Inhomogeneous Media (Ohio State University)}, Proceeding of the first conference on the Electrical Transport and Optical Properties of Inhomogeneous Media, AIP Conf. Proc. No. 40, edited by Garland J. C. and Tanner D. B. 1978 (AIP, New York).
\bibitem{02} Bergman D J and Stroud D 1992 {\it Solid state Physics} {\bfseries 46} 147.
Clerc J P, Giraud G, Laugier J M and Luck J M, 1990 {\it Adv. Phys.} {\bfseries 39}, 191.
\bibitem{03} Stauffer D. and Aharony A., {\it Introduction to Percolation Theory, 2nd Ed. (Taylor and} Francis, London, 1994).
\bibitem{04} Dykhne A M 1970 {\it Zh.Eksp. Teor. Fiz.} {\bfseries 59}, 110(Engl.Transl.1971 {\it Sov. Phys.-JETP} {\bfseries 32}348).
Bruggeman D A 1935 {\it Ann. Phys.,Leipzig.} {\bfseries 24} 636.
\bibitem{05} Landauer R 1952 {\it J. Appl. Phys.} {\bfseries 23}, 779
\bibitem{06} Zeng X. C., Hui P. M. and Stroud D. 1989 {\it Phys. Rev B} {\bfseries 39}, 1063.
\bibitem{07} Koss R. S. and Stroud D. 1987 {\it Phys. Rev B} {\bfseries 35}, 9004.
\bibitem{08} Waston B. P. and Leath P. 1974 {\it Phys. Rev B} {\bfseries 9}, 4893.
\bibitem{09} Cummings K. D., Garland J. C. and Tanner D. B. 1984 {\it Phys. Rev B} {\bfseries 30}, 4170.
\bibitem{10} Brouers F., Sarychev A. K., Blacher S. and Lothaire O. 1997 {\it Physica. A} {\bfseries 241}, 146.
\bibitem{11} Zekri L, Zekri N, Bouamrane R and Brouers F 2000 {\it J. Phys.:Condens. Matt.} {\bfseries 12} 293.
\bibitem{12} Lobb C. J. and Frank D. J. 1984 {\it Phys. Rev B} {\bfseries 30}, 4090; Frank D. J. and Lobb C. J. 1988 {\it Phys. Rev B} {\bfseries 37}, 302.
\bibitem{13} Zekri L, Bouamrane R and Zekri N 2000 {\it J.Phys.A:Math.Gen.} {\bfseries 33} 649.
\bibitem{14} Kreibig U and Vollmer M, {\it Optical Properties of Metal Clusters} (Springer-Verlag, Berlin, 1995).
 Wooten F, {\it Optical Properties of Solids} (Academic, New York, 1972)
\bibitem{15} Clerc J P, Zekri L and Zekri N submitted to {\it Phys. Lett. A}
\bibitem{16} Berhier S, {\it Optique des milieux composites} 1993 Polytechnica, Paris

\end{thebibliography}
\end{document}